\documentclass[aps,prl,twocolumn,floats,showpacs]{revtex4}
\usepackage{epsfig}
\usepackage{color}
\usepackage{bm}
\usepackage{latexsym}

\begin{document}
\newcommand{\const}{\mbox{const}}
\newcommand{\eexp}{\mbox{e}^}
\newcommand{\bra}{\left\langle}
\newcommand{\ket}{\right\rangle}

\title{Current relaxation in nonlinear random media}
\author{Tsampikos Kottos$^{1}$ and Matthias Weiss$^{2}$}
\affiliation{$^1$ Max-Planck-Institut f\"ur Str\"omungsforschung, Bunsenstra\ss e 10,
D-37073 G\"ottingen, Germany}
\affiliation{$^{2}$ MEMPHYS-Center for Biomembrane Physics, Physics Dept.,
  University of Southern Denmark, Campusvej~55, DK-5230 Odense M, Denmark}


\begin{abstract}
We study the current relaxation of a wave packet in a nonlinear random sample 
coupled to the continuum and show that the survival probability decays as $P(t) \sim 1/t^{\alpha}$. 
For intermediate times $t<t^*$, the exponent $\alpha$ satisfies a scaling law 
$\alpha =f(\Lambda=\chi/l_{\infty})$ where $\chi$ is the nonlinearity strength and $l_{\infty}$ 
is the localization length of the corresponding random system with $\chi=0$. For $t\gg t^*$ 
and $\chi>\chi_{\rm cr}$ we find a universal decay with $\alpha=2/3$ which is a signature of the 
{\it nonlinearity-induced delocalization}. Experimental evidence should be observable in coupled 
nonlinear optical waveguides.
\\
\end{abstract}

\maketitle

A fundamental source of physical information are time-resolved decay measurements in quantum 
mechanical systems which are coupled to a continuum via metallic leads or absorbing boundaries. 
While the radioactive decay is a prominent paradigm, more recent examples include atoms in 
optically generated lattices and billiards \cite{RSN97,F01}, the ionization of molecular 
Rydberg states \cite{B91}, photoluminescence spectroscopy of excitation relaxation in 
semiconductor quantum dots and wires \cite{B99}, and pulse propagation studies with 
electromagnetic waves \cite{CZG03}. Motivated by the experimental achievements, the dynamics 
of open quantum systems has also gained considerable interest from a theoretical perspective 
and various analytical techniques have been developed to study the problem in more detail 
\cite{AKL87,SS97,F97,M00,CMS00,CGM00,BCMS00,OWKG01}. One possible approach to the problem is 
to consider the survival probability $P(t)$ of a wave packet which is initially localized 
inside an open sample of size $N$. The total current leaking out of the sample is then related 
to the survival probability by $J(t)=-{\partial P(t) \over \partial t}$.

For ballistic/chaotic systems \cite{SS97}, and for random systems in the metallic
regime \cite{M00}, $P(t)$ is by now well understood. Recently, also quantum systems 
with a mixed classical phase space have been studied \cite{CMS00,BCMS00}, where it 
was found that $P(t)\sim 1/t$. The same algebraic decay was found \cite{CMS00} for 
disordered (or dynamical) systems with exponential localization \cite{A56,I90,OKG02}. 
In both cases, this law is related to localization and tunneling effects, and applies 
for intermediate asymptotic times \cite{CMS00}. For larger times, localization effects 
lead to a log-normal decay of the survival probability \cite{M00}. 

The subject of the present paper is the survival probability in a new setting, namely a 
class of {\it random} systems where the evolution is governed by a {\it Discrete Nonlinear
Schr\"odinger Equation} (DNLSE) (see \cite{HT99} and references therein), that is
\begin{equation}
\label{dnls}
i{\partial \over \partial t} \psi_n(t) = \sum_m H_{nm} \psi_m(t)
- \chi |\psi_n(t)|^2 \psi_n(t) \,\,.
\end{equation}
Here, $|\psi_n(t)|^2$ denotes the probability for a particle to be at site $n$ at time 
$t$, $H_{nm}$ is a random tight-binding Hamiltonian and $\chi$ is the strength of the 
nonlinearity. The nonlinear term in Eq.~(\ref{dnls}) can arise due to a mean field 
approximation for many-body interactions, e.g. in the Gross-Pitaevskii framework for 
Bose-Einstein Condensates \cite{DGPS99}, or it can result from the description of a 
quantum mechanical particle which moves in a random 
potential and interacts strongly with vibrations \cite{KC86}. The DNLSE can be also 
viewed as a description for the energy transfer in proteins \cite{D73}. Finally, in 
the context of optics the DNLSE is capable of describing wave motion in coupled nonlinear 
optical waveguides \cite{CJ88,BCPS91,S93}. In the latter case the longitudinal space 
dimension of the waveguide plays the role of the time variable.

Here, we present for the first time the consequences on the decay of the survival
probability $P(t)$, i.e. on the flux out of the sample when coupling a nonlinear random system 
like Eq.~(\ref{dnls}) to the continuum via absorbing boundaries (or conducting leads).
In particular, we find that for intermediate (but large) times
\begin{equation}
\label{powlaw}
P(t) \approx {C\over t^{\alpha}}\,\quad{\rm with} \quad \alpha(\chi,l_{\infty})=f(\Lambda={\chi\over l_{\infty}})\,
\end{equation}
where $C$ is some constant and $f(\Lambda)$ is a universal scaling function which encodes
the interplay between the nonlinearity and the disorder. We find that $f(\Lambda<\Lambda^*)
\approx 1$ while $f(\Lambda\gg\Lambda^*) \approx 0.35\pm0.05$ where $\Lambda^*\sim {\cal O}(1)$.

Moreover, for nonlinearity strengths $\chi$ above the {\it delocalization border} $\chi_{\rm cr}$ 
at which localization phenomena are destroyed \cite{S93}, and for times $t>t^*\sim N^{2/\nu}/
(\chi l_{\infty})^2$ (where $\nu \approx 2/5$ is determined by the asymptotic spreading of the 
wavepacket's variance ${\rm var}(t) \sim t^{\nu}$), we find a universal asymptotic power-law decay 
\begin{equation} \label{powlaw1}
P(t)\approx C\left({\chi\over t}\right)^{2/3}. 
\end{equation}
The intermediate power law (2) is observed for a large class of localized initial excitations (i.e. 
$\delta$-like, narrow Gaussians) near the surface of the sample which have attracted both 
experimental and theoretical interest \cite{CZG03,CMS00,CGM00,OWKG01,BCMS00} in recent years. 
On the other hand (3) is independent of the initial condition and
applies as soon as the wavepacket is spread ergodically over the
sample. Our results (\ref{powlaw},\ref{powlaw1}) are confirmed 
numerically and are supported by theoretical arguments.
\begin{figure}
\begin{center}
    \epsfxsize=8.4cm
    \leavevmode
    \epsffile{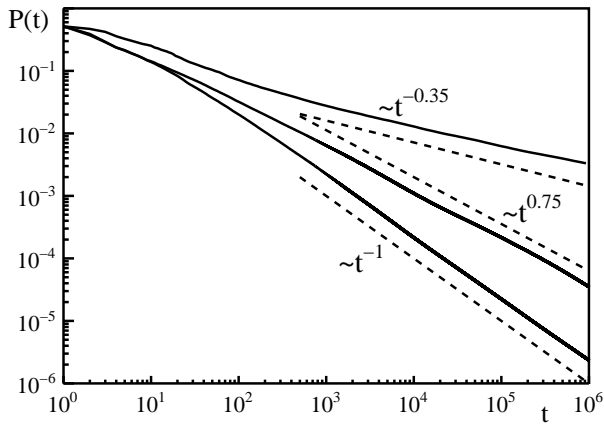}
\caption{The survival probability $P(t)$ [Eq.~(\ref{geom})] shows a power-law decay 
$P(t)\sim t^{-\alpha}$ for intermediate (but large) times $t<t^*$. Shown are three 
representative examples for $k=5,\,K=7$, $N=1024$ and $\chi=1,\,15,\,200$ (bottom to top). 
}
\label{fig:fig1}
\end{center}
\end{figure}

To investigate the decay of the survival probability we use a modified version \cite{S93} of the 
Kicked Rotor (KR) with absorbing boundary conditions \cite{CMS00}. Based on the similarities 
\cite{I90,OKG02} between dynamical and Anderson localization, it is expected \cite{S93,M98} that 
the same dynamics will be generated by Eq.~(\ref{dnls}). The discrete quantum mechanical evolution 
from time $t$ to time $t+1$ (measured in units of a kick period $T$) is described by the map 
\begin{equation}
\psi_n(t+1) = \sum_m e^{i\pi(m-n)} J_{n-m}(k) e^{-iT((m+\phi)^2-\chi |\psi_m|^2)} \psi_m(t)
\label{nlkr}
\end{equation}
where the Bessel function $J_{n-m}$ appears as the result of the kick described by the operator
$U_{\rm kick}= \exp(-ik\cos(\theta))$. Here, $\theta$ denotes the angle, $n$ is the angular 
momentum while $K=kT$ is the classical kicking strength. The parameter $\phi$ can be interpreted 
as an Aharonov-Bohm flux through the ring parameterized by the coordinate $\theta$. In contrast 
to the standard KR now the kinetic term depends on the wavefunction probability. The modified
KR ~(\ref{nlkr}) models propagation of nonlinear waves in optical fibers with a change of the 
optical density inside the waveguide \cite{BCPS91,S93}. The same model approximately describes 
propagation in waveguides with longitudinal sinusoidal modulation of the boundary \cite{BCPS91,S93}. 

The dynamics of the model (\ref{nlkr}) for $\chi=0$ is by now well studied. The classical motion 
is chaotic and for sufficiently large values of $K$ there is diffusion in momentum space with 
diffusion coefficient $D\simeq {k^2\over 2}$ \cite{I90}. The most striking consequence of 
quantization is the suppression of this classical diffusion due to quantum-dynamical localization 
\cite{I90}, the dynamical version of the well-known Anderson localization \cite{A56}. The 
eigenstates of the associated unitary operator $\cal U$ are exponentially localized with a 
localization length $l_{\infty} \simeq k^2$. The dynamical localization, however, was found to 
be destroyed for nonlinearity strengths $\chi>\chi_{\rm cr}\sim 1/T$ \cite{S93}. A subdiffusive 
growth ${\rm var}(t)\approx (T \chi l_{\infty})^{4/5} t^{\nu}$ of the wave packet's variance 
emerged, leading to a uniform spreading over the whole sample. The value $\nu=2/5$ was deduced in 
\cite{S93} on the basis of the Chirikov criterion of overlapping resonances and was found to be
in reasonable agreement with numerical results. For times $t>t^*\sim {N^5\over(T \chi l_{\infty})^2}$ 
the subdiffusion leads to a uniform distribution of an initially localized wavepacket over the 
sample of size $N$. 
\begin{figure}
\begin{center}
    \epsfxsize=8.4cm
    \leavevmode
    \epsffile{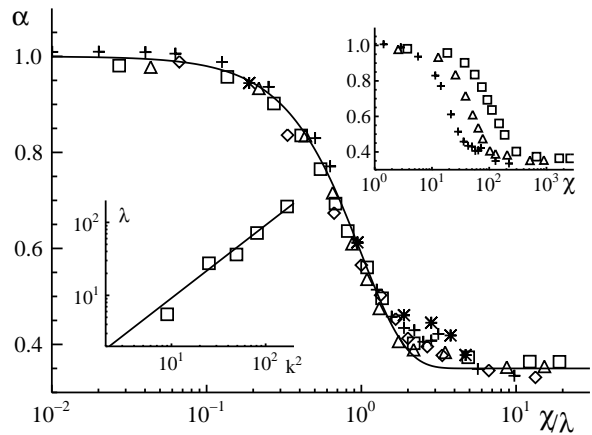}
\caption{
Power-law exponents~$\alpha$ of the survival probability as a function of the scaled 
nonlinearity strength~$\chi/\lambda$ for various localization lengths $l_{\infty}\approx 
k^2$ with a heuristic scaling function (full line). Error bars are smaller than the 
symbol size. Lower inset: The scaling parameter $\lambda$ increases linearly with 
$l_{\infty}$ (full line, slope $\approx 1$). Upper inset: Power-law exponents~$\alpha$ 
vs. the nonlinearity strength~$\chi$ for $k=5 (+)\,,9(\bigtriangleup)\,,13(\Box)$.
}
\label{fig:fig2}
\end{center}
\end{figure}

To open the system described by Eq.~(\ref{nlkr}) we additionally apply a projection operator 
${\cal P}$ to the wavepacket in a fixed interval of momentum states $0\leq n\leq N$, i.e. 
$|{\tilde \psi}(t)\rangle={\cal{P}}|\psi(t)\rangle$. Thus, ${\cal P}$ describes the complete deletion 
(absorption) of the part of the wavepacket which propagates outside the given interval. 
The survival probability inside the sample at a time $t$ is then given by 
$p(t) \equiv |{\tilde{\psi}}(t)|^{2} = \langle{\tilde \psi}(t)|{\tilde{\psi}}(t)\rangle$.
To suppress fluctuations, we concentrate on the geometric average of $p(t)$, i.e.
\begin{equation}\label{geom}
P(t)=\exp(\langle \ln(p(t))\rangle_{\phi}),
\end{equation}
where $\langle \cdot \rangle_{\phi}$ indicates an averaging over different phases $\phi$ 
(typically more than $20$). The initial excitation (unless stated otherwise) is a $\delta$-like wave packet launched 
at one of the boundaries, i.e. $\psi_n(t=0)=\delta_{n,0}$. In our numerical calculations, 
we have used $K=kT=7$ and $k = 3,5,7,9,13$ while the sample length $N$ was chosen to fulfill 
always $N\gg l_{\infty}$. Due to localization, the survival probability shows a decay $P(t)
\sim 1/t$ for $\chi=0$ \cite{CMS00}.

In Fig.~\ref{fig:fig1} we report the survival probability $P(t)$ for $k=5$, $N=1024$ and three 
representative values of the nonlinear coupling ($\chi=1,\,15,\,200$). In all cases, $P(t)$ 
clearly displays a power-law decay $P(t)\sim t^{-\alpha}$ for times $t<t^*$. The exponents 
$\alpha$ for various nonlinearities $\chi$ and localization lengths $l_{\infty}$ are summarized 
in Fig.~\ref{fig:fig2}. All curves have the same functional form, albeit being shifted with 
respect to each other (Fig.~\ref{fig:fig2}, upper inset). The curves $\alpha(\chi)$ do coincide, 
however, when plotting them versus $\Lambda=\chi/\lambda$ where $\lambda$ is a scaling parameter 
(Fig.~\ref{fig:fig2}, main part). We find that the scaling parameter~$\lambda$ depends linearly 
on $l_{\infty}$ (Fig.~\ref{fig:fig2}, lower inset) resulting in a one-parameter scaling of the 
power-law exponent according to Eq.~(\ref{powlaw}). Our data indicate that for $\Lambda<\Lambda^*
\approx 0.5$ localization effects are dominant and the survival probability decays as $P(t)\sim 
1/t$. In the opposite limit of large nonlinearities the decay of $P(t)$ follows a universal law 
$P(t)\sim 1/t^{\alpha}$ with $\alpha\approx0.35\pm0.05$. We found that the smooth transition between 
the two limits is characterized by the scaling function $f(\Lambda)=0.35+0.65\exp(-\Lambda^{3/2})$.
\begin{figure}
\begin{center}
    \epsfxsize=8.4cm
    \leavevmode
    \epsffile{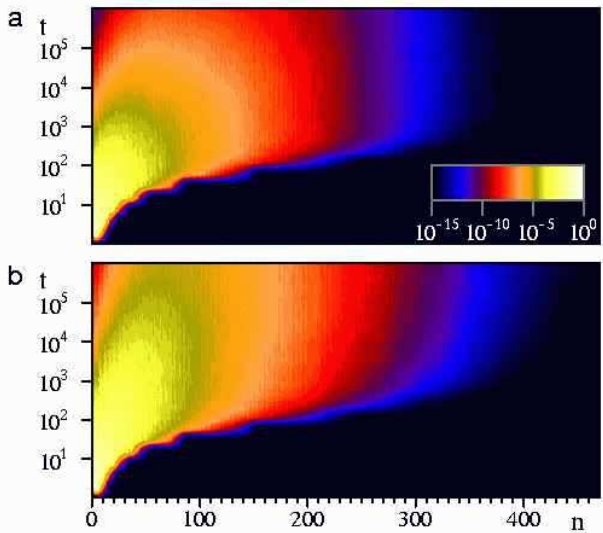}
\caption{
Density plots of a wave-packet evolution for (a) $\chi=1$ and (b) $\chi=50$ (parameters as in 
Fig.~\ref{fig:fig1}). Notice the development of probability depletion near the absorbing boundary for 
$\chi=50$ which leads to the creation of a potential barrier $V_{\chi}= -\chi |\psi_n(t)|^2$. 
The time axis and the color-coded height of the wave function (see inset in a) are both on a 
log-scale.}
\label{fig:fig3}
\end{center}
\end{figure}

In Fig.~\ref{fig:fig3} we report the spatio-temporal evolution of the wave packet for two 
limiting values of $\chi$ in a density plot. For $\chi=1$ the initial excitation 
is essentially localized in a region close to the absorbing boundary, covering 
the interval $0\leq n\leq l_{\infty}$ in a uniform manner (Fig.~\ref{fig:fig2}a). This is 
the same behavior as found for $\chi=0$ (data not shown). As the nonlinearity increases, 
subdiffusion takes over, leading to a spreading of the dominant fraction of the wavefunction
into the bulk of the sample. This results in a gradual depletion of the wavefunction from 
the boundary zone and a creation of an effective potential barrier, the height of which 
increases with the nonlinearity strength $\chi$ as $V_{\chi} (t)=-\chi |\psi_n(t)|^2$. 
This barrier traps the probability inside the bulk of the sample and obstructs the outwards 
flux thus leading to the observed slow decay of $P(t)$ for intermediate times $t< t^*$ 
\cite{note1}. 
\begin{figure}
\begin{center}
    \epsfxsize=8.4cm
    \leavevmode
    \epsffile{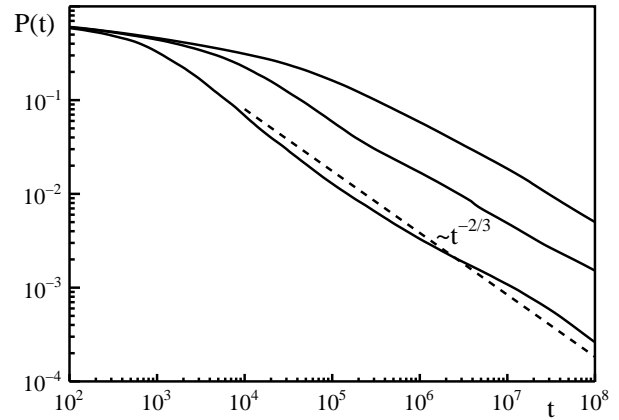}
\caption{Decay of $P(t)$ for $N=256$, $k=5$, $K=7$ and $\chi=2,\,10,\,50$ (from top); 
the numerical result is in agreement with Eq.~(\ref{powlaw1}) (dashed line). We used 
a uniform (ergodic-like) excitation as initial condition.
}
\label{fig:fig4}
\end{center}
\end{figure}

We want to understand the one-parameter scaling of the power-law exponent $\alpha
(\chi,l_{\infty})=f(\Lambda)$ (see Fig.~\ref{fig:fig2}). We recall that the slow 
decay to the continuum in the presence of the nonlinearity is due to tunneling through a 
barrier of height $V_{\chi}\sim \chi |\psi|^2$ which is created in a regime $n \sim l_{
\infty}$ from the boundary. Assuming that for very small nonlinearities the components 
$\psi_n$ of the wave packet are similar to the case $\chi=0$, i.e. $|\psi_n|\sim \exp(-n
/l_{\infty})/\sqrt{l_{\infty}}$, we obtain $V_{\chi} \sim \chi/l_{\infty}$. 
The barrier is totally transparent if $V_{\chi} \leq E $ where 
$E\sim {\cal O} (1)$ is the energy of the excitation. Consequently we get that $\Lambda^* 
\equiv\chi^*/l_{\infty}\sim {\cal O}(1)$ indicating that the ratio $\Lambda = \chi/l_{\infty}$ 
controls the behavior of $P(t)$ and therefore the value of $\alpha$ (see Eq.~(\ref{powlaw})). 

Next, we give a qualitative argument for the power-law decay~(\ref{powlaw}) when $\Lambda
\lesssim \Lambda^{*}$. We start with the perturbative expression for the survival probability 
for $\chi=0$ \cite{CMS00}, i.e.
\begin{equation}\label{pt0}
P(t)\sim \int_0^{\infty} |c_k|^2 e^{-\Gamma_k t} dk\,\,,
\end{equation}
where $|c_k|^2=|\langle\psi(t=0)|u_k\rangle|^2\sim\exp(-2k/l_{\infty})/l_{\infty}$ are the 
overlaps of the initial state $|\psi(t=0)\rangle$ with the exponentially localized eigenstates 
$|u_k\rangle$ of the operator~$\cal U$ [Eq.~(\ref{nlkr})] for $\chi=0$. The decay $P(t)\sim1/t$ 
results from Eq.~(\ref{pt0}) when using $\Gamma_k\sim T_k\sim \exp(-2k/l_{\infty})/l_{\infty}$, 
where $T_k$ is the transmission probability of the $k-$th mode. In the presence of a nonlinearity,
it is known that the transmission probability still decays exponentially \cite{M98,DR87}, and 
therefore it is reasonable to assume that $T_k\sim \exp(-2k/l_{\chi})$ where $l_{\chi} = \alpha 
l_{\infty}$ is an effective length determined by both, the tunneling through the barrier 
$V_{\chi}$ and the disordered potential. Since the existence of the barrier additionally hampers 
the transmission $T_k$ with respect to $\chi=0$, it is reasonable to assume that $\alpha <1$. 
Substituting this expression for $T_k$ into Eq.~(\ref{pt0}) we obtain Eq.~(\ref{powlaw}).

For times $t\geq t^*$ and $\chi>\chi_{\rm cr}$, the subdiffusion leads to an ergodic-like 
distribution of the initial excitation over the whole sample \cite{S93}. A simple approach 
based on a modified diffusion equation suggests that $P(t)$ should consequently decay in a 
stretched-exponential manner \cite{note2}. It is thus surprising that our data contradict 
this very expectation (see Fig.~\ref{fig:fig4}). In fact, we find a {\it universal power-law} 
decay with exponent $\alpha \approx 2/3$. This can be understood by working in the basis of 
localized eigenstates $|u_k\rangle$ of the linear case ($\chi=0$). In this representation,
we can estimate the transition rate from a quasi-energy level $E_k$ to
other levels in a discance $\sqrt{{\rm var}(t)}$ to be 
$\Gamma_k \sim (T\chi)^2/[{\rm var}(t)]^{3/2}$ \cite{S93}.
Thinking semi-classically, the latter can be associated with the escape time 
$t_k\sim \Gamma_k^{-1}$ of a particle that was initially located at a distance 
$\sqrt{{\rm var}(t)}$ from the absorbing boundary. At any time $t$ the number $\mu$ 
of particles with inverse escape times $\Gamma_k> t^{-1}$ is thus given by
$\mu \sim \sqrt{{\rm var}(t)}\sim [(T\chi)^2 t]^{1/3}$. As the probability to survive 
inside the sample up to a time $t$ is given by $P(t) \propto d\mu/dt$ \cite{CMS00}, this
leads to $P(t) \sim (T\chi/t)^{2/3}$ in agreement with our data (see Fig.~\ref{fig:fig4}).

We have checked also that for $\chi<\chi_{\rm cr}$ localization effects are dominant leading to 
a decay of $P(t)$ as in the case of $\chi=0$ \cite{M00,CMS00}. Finally, in the limit 
$t\to\infty$, $|\psi_n(t)|^2$ becomes very small due to the loss of norm from the absorbing 
boundaries and $P(t)$ decay as for $\chi=0$ \cite{M00,CMS00}. 

We thank D.~Cohen for helpful comments. This research was supported by the German-
Israeli Foundation for Scientific Research and Development (GIF). The MEMPHYS-Center 
for Biomembrane Physics is supported by the Danish National Research Foundation.


\begin{thebibliography}{99}

\bibitem{RSN97}M. Raizen, C. Salomon, Q. Niu, Phys. Today {\bf 50}, 30 (1997);
D. S. Wiersma et al., Nature {\bf 390}, 671 (1997); K. W. Madison, et al., 
Phys. Rev. A {\bf 60}, R1767 (1999).

\bibitem{F01} N. Friedman et al., Phys. Rev. Lett. {\bf 86}, 1518 (2001); A. Kaplan
et al., ibid. {\bf 87}, 274101 (2001); V. Milner et al., ibid. {\bf 86}, 1514 (2001).

\bibitem{B91} T. Baumert et al., Phys. Rev. Lett. {\bf 67}, 3753 (1991).

\bibitem{B99} G. Bacher et al., Phys. Rev. Lett. {\bf 83}, 4417 (1999); R. Kumar et al.,
ibid. {\bf 81}, 2578 (1998).

\bibitem{CZG03}A. A. Chabanov, et al., Phys. Rev. Lett. {\bf 90}, 203903 
(2003); S. E. Skipetrov and B. A. van Tiggelen, ibid. {\bf 92}, 113901 (2004); L. Dal Negro 
et al., ibid. {\bf 90}, 055501 (2003); A. Z. Genack, et al., ibid., {\bf 82} 715 (1999). 

\bibitem{AKL87} B. L. Altshuler, V. E. Kravtsov, I. V. Lerner, Pisma Zh.
Eksp. Teor. Fiz. {\bf 45}, 160 (1987) [ JETP Lett. {\bf 45}, 199 (1987)];
B. L. Altshuler, V. N. Prigodin, Zh. Eksp. Teor. Fiz. {\bf 95}, 348 (1989) 
[ Sov. Phys. JETP {\bf 68}, 409 (1980)]

\bibitem{SS97} D. V. Savin and V. V. Sokolov, Phys. Rev. E {\bf 56}, R4911
(1997); F. M. Dittes, Phys. Rep. {\bf 339}, 215 (2000); M. Puhlmann et al.,
nlin.CD/0401037.


\bibitem{F97}K. M. Frahm, Phys. Rev. E {\bf 56}, 6237 (1997); G. Casati, G. Maspero,
D. L. Shepelyansky, ibid. {\bf 56}, 6233 (1997).

\bibitem{M00} A. D. Mirlin, Phys. Rep. {\bf 326}, 259 (2000); B. A. Muzykantskii and 
D. E. Khmelnitskii, Phys. Rev. B {\bf 51}, 5480 (1995); I. E. Smolyarenko and B. L. 
Altshuler, Phys. Rev. B {\bf 55}, 10451 (1997).

\bibitem{CMS00} G. Casati, G. Maspero, and D. L. Shepelyansky, Phys. Rev Lett.
{\bf 82}, 524 (1999); ibid. {\bf 84}, 4088 (2000); G. Benenti, et al., ibid.
{\bf 87}, 014101 (2001).


\bibitem{CGM00} G. Casati, I. Guarneri, and G. Maspero, Phys. Rev Lett.
{\bf 84}, 63 (2000); L. Hufnagel, R. Ketzmerick, and M. Weiss, 
Europhys. Lett. {\bf 54}, 703 (2001).

\bibitem{OWKG01} A. Ossipov, et al., Phys. Rev. B, {\bf 64} 224210, (2001).

\bibitem{BCMS00}G. Benenti et al., Phys. Rev. Lett. {\bf 84}, 4088 (2000); S. Wimberger, 
et al., ibid. {\bf 89}, 263601 (2002).

\bibitem{A56} P. W. Anderson, Phys. Rev. {\bf 109}, 1492 (1958); A. MacKinnon and
B. Kramer, Rep. Prog. Phys. {\bf 56}, 1469 (1993); E. Abrahams, et al., Phys. Rev.  
Lett. {\bf 42}, 673 (1979);
L. P. Gorkov, A. I. Larkin, and D. E. Khmelnitskii, JETP Lett. {\bf 30} 228 (1979).

\bibitem{I90} F. M. Izrailev, Phys. Rep. {\bf 196}, 300 (1990); G.~Casati et al., Lect. 
Notes Phys. {\bf 93}, 334 (1979).

\bibitem{OKG02} A. Ossipov, T. Kottos and T. Geisel, Phys. Rev E {\bf 65}, 055209(R) (2002);
S. Fishman, D. R. Grempel, R. E. Prange, Phys. Rev. Lett. {\bf 49}, 509 (1982); M. Weiss, 
T. Kottos and T. Geisel, Phys. Rev. B, {\bf 63} 081306(R) (2001). 

\bibitem{HT99} D. Hennig, G. P. Tsironis, Phys. Rep. {\bf 307}, 333 (1999).

\bibitem{DGPS99} F. Dalfovo et al., Rev. Mod. Phys. {\bf 71}, 463 (1999).

\bibitem{KC86} V. Kenkre, D. Campbell, Phys. Rev. B {\bf 34}, 4959 (1986).

\bibitem{D73} A. S. Davydov, J. Theor. Biol. {\bf 38}, 559 (1973); A. S. Davidov, 
N. I. Kislukha, Sov. Phys., JETP {\bf 44}, 571 (1976); Phys. Status. Sol. B {\bf 59}, 
465 (1973).

\bibitem{CJ88} D. N. Christodoulides, R. I. Joseph, Opt. Lett. {\bf 13}, 794 (1988);
Phys. Rev. Lett. {\bf 62}, 1746 (1989); D. Mandelik, H. S. Eisenberg, Y. Silberberg,
R. Morandott, J. S. Aitchinson, ibid. {\bf 90}, 253902 (2003); ibid {\bf 90}, 053902
(2003).

\bibitem{BCPS91} F.~Benvenuto, et al., Phys. Rev. A {\bf 44}, R3432 (1991).

\bibitem{S93} D.L.~Shepelyansky, Phys. Rev. Lett. {\bf 70}, 1787 (1993).

\bibitem{M98} M. I. Molina, Phys. Rev. B {\bf 58}, 12547 (1998); J. C. Eilbeck and M. Johansson,
nlin.PS/0211049; G. P. Tsironis, in {\it Dynamical Studies of the Discrete Non-linear Schr\"odinger
Equation}, Erasmus Lectures (1994).

\bibitem{note1} 
We confirmed numerically that the exponent~$\nu$ of ${\rm var}(t)\sim t^\nu$ 
is independent of $\chi$ on time scales on which we observe Eq.~(\ref{powlaw});
this also was true for the exponent of the return probability $P_s(t)=|\bra\psi_0|\psi(t)\ket|^2$.
Therefore, subdiffusion alone cannot explain the dependence $\alpha(\chi)$.

\bibitem{DR87} B. Doucot, R. Rammal, Europhys. Lett. {\bf 3}, 969 (1987).

\bibitem{note2} Following \cite{CZG03}, we can use a diffusion equation with a time-dependent 
diffusion coefficient and absorbing boundaries to obtain $P(t) \sim \exp(-\int_0^t D(t') dt')$. 
For $D(t)=D_0$, we obtain an exponential decay of $P(t)$; in our case $D(t)\sim t^{-3/5}$ 
leads to $P(t)\sim \exp(-t^{2/5})$. This approach can also take localization corrections into 
account \cite{CZG03,F97,M00}. 
\end{thebibliography}
\end{document}